\documentclass[12pt]{amsart}

\usepackage[centertags]{amsmath}
\usepackage{amsthm,amsfonts,amssymb}
\usepackage[T1]{fontenc}
\usepackage{lmodern,enumitem,indentfirst,cite}
\theoremstyle{plain}
\newtheorem{theorem}{Theorem}[section]
\newtheorem{lemma}[theorem]{Lemma}

\theoremstyle{definition}
\newtheorem{remark}[theorem]{Remark}
\newcommand{\F}{\mathbb F}
\newcommand{\C}{\mathcal C}
\newcommand{\GL}{\operatorname{GL}}
\newcommand{\PGL}{\operatorname{PGL}}
\newcommand{\Span}{\operatorname{Span}}
\newcommand{\Tr}{\operatorname{Tr}}
\title[Linear equivalence of FWS cyclic orbit codes]{Linear equivalence of full weight spectrum cyclic orbit codes}

\author[Minjia Shi]{Minjia Shi}
\address{School of Mathematical Sciences, Anhui University, Hefei 230601, China}
\email{smjwcl.good@163.com}
\author[Ferruh \"Ozbudak]{Ferruh \"Ozbudak}
\address{Faculty of Engineering and Natural Sciences, Sabanc{\i} University, Istanbul, T\"urkiye}
\email{ferruh.ozbudak@sabanciuniv.edu}
\date{3 November 2026}
\subjclass[2020]{94B27, 11T71}
\keywords{Cyclic subspace code, full weight spectrum, linear equivalence, finite field}

\begin{document}
\pagestyle{plain}
\begin{abstract}
We determine the linear-equivalence classes within the two families of full weight spectrum one-orbit cyclic subspace codes. This solves the second open problem posed by Castello, Polverino and Zullo [J. Combin. Theory Ser. A 212 (2025) 106005]. For the first family, the criterion is a fractional linear transformation over $\F_q$ followed by a power of $q$-Frobenius. For the second family, it is an affine transformation over $\F_{q^2}$ together with a power of $q$-Frobenius. We use polynomial spaces over finite fields, the linear-isometry theorem for cyclic orbit codes, and trace duality. The arguments include parameters belonging to proper subfields of the ambient field.
\end{abstract}
\maketitle

\section{Introduction and main result}

Subspace codes were introduced by Koetter and Kschischang~\cite{KK} for error correction in random linear network coding. Cyclic codes in projective space were introduced by Etzion and Vardy~\cite{EV}, and orbit codes under cyclic subgroups of the general linear group were studied by Trautmann, Manganiello, Braun and Rosenthal~\cite{TMBR}. Unions of cyclic orbit codes with large minimum distance were constructed by means of subspace polynomials in \cite{BEGR,OO1,OO2}.

Let $q$ be a prime power, let $n\geq2$ be an integer, and put $L=\F_{q^n}$. For an $\F_q$-subspace $U$ of $L$, write
\[
 \C_L(U)=\{aU:a\in L^*\}.
\]
This is the one-orbit cyclic subspace code, or cyclic orbit code, defined by $U$. We say that $\C_L(U)$ and $\C_L(V)$ are \emph{linearly equivalent} if
\[
 T\bigl(\C_L(U)\bigr)=\C_L(V)
 \qquad\text{for some }T\in\GL_{\F_q}(L).
\]
Thus the equivalence is induced by an invertible $\F_q$-linear map on the ambient field. In \cite{CPZ,GLL} linearly equivalent codes are called linearly isometric. For two $k$-dimensional subspaces, the subspace distance is
\[
 d(U,V)=2k-2\dim_{\F_q}(U\cap V).
\]
A one-orbit cyclic subspace code has \emph{full weight spectrum} (FWS) if all the distances $2,4,\ldots,2k$ occur.

Roth, Raviv and Tamo~\cite{RRT} observed that, for $k=\dim_{\F_q}U\geq2$, the code $\C_L(U)$ has $(q^n-1)/(q-1)$ elements and minimum distance $2k-2$ if and only if $U$ is a Sidon space. The equivalence of Sidon spaces was studied in~\cite{CPSZ}. Multi-orbit cyclic subspace codes were related to linear sets in~\cite{Zu} and used to construct optical orthogonal codes in~\cite{OSZ}.

Since $d(aU,bU)=d(U,a^{-1}bU)$ for $a,b\in L^*$, the distance distribution of $\C_L(U)$ is determined by the dimensions of the intersections $U\cap aU$ with $a\in L^*$. Gluesing-Luerssen and Lehmann~\cite{GL} studied this distribution, which is a finer invariant of cyclic orbit codes than the minimum distance. If $\dim_{\F_q}U=k$, the distance between two distinct codewords lies in $\{2,4,\ldots,2k\}$, and so $\C_L(U)$ has at most $k$ distinct nonzero distances. The FWS codes are those attaining this bound.

The analogous question in the Hamming metric is classical; see Delsarte~\cite{De73}. Shi, Zhu, Sol\'e and Cohen~\cite{SZSC} proved that a linear code of dimension $m$ over $\F_q$ has at most $(q^m-1)/(q-1)$ distinct nonzero weights, and codes attaining this bound exist for all $q$ and $m$ by~\cite{AN,SZSC}. A linear code of length $N$ has at most $N$ distinct nonzero weights, and Alderson~\cite{Al} determined the parameters of the codes attaining this bound, which are called full weight spectrum codes. The corresponding question for cyclic codes was studied by Shi, Li, Neri and Sol\'e~\cite{SLNS}.

Castello, Polverino and Zullo \cite[Theorem 1.2]{CPZ} classified the FWS one-orbit cyclic subspace codes into the following two families. When $k=1$, the orbit is the set of all one-dimensional $\F_q$-subspaces of $L$, so there is only one code. We therefore assume $k\geq2$.

\smallskip
\noindent\textbf{Family I.} For $\lambda\in L\setminus\F_q$, put
\[
 U_k(\lambda)=\Span_{\F_q}\{1,\lambda,\ldots,\lambda^{k-1}\},
 \qquad t_\lambda=[\F_q(\lambda):\F_q].
\]
The admissible range is
\begin{equation}\label{eq:range-I}
 \begin{cases}
  2k\leq n,&\text{if }t_\lambda=n,\\
  2k-1\leq t_\lambda,&\text{if }t_\lambda<n.
 \end{cases}
\end{equation}

\smallskip
\noindent\textbf{Family II.} Suppose $n$ is even and put $K=\F_{q^2}\subseteq L$. Let $\lambda\in L\setminus K$ and $t_\lambda=[K(\lambda):K]$. For an integer $r$ with $1\leq r\leq t_\lambda$, define
\[
 A_r(\lambda)=\Span_K\{1,\lambda,\ldots,\lambda^{r-1}\}.
\]
For a positive integer $\ell$, put
\[
 S_\ell(\lambda)=A_\ell(\lambda)\oplus\F_q\lambda^\ell.
\]
Here the admissible range is
\begin{equation}\label{eq:range-II}
 t_\lambda>2\ell,\qquad k=2\ell+1.
\end{equation}
In particular, $\dim_{\F_q}S_\ell(\lambda)=2\ell+1$ and $[K(\lambda):\F_q]=2t_\lambda\geq2k$.

Linear and semilinear isometries of subspace codes were studied by Trautmann~\cite{Tr}. For cyclic orbit codes, Gluesing-Luerssen and Lehmann~\cite{GLL} proved that if $1\in U$, $2\leq\dim_{\F_q}U\leq n/2$ and $U$ is not contained in a proper subfield of $L$, then every linear isometry from another cyclic orbit code onto $\C_L(U)$ is of the form $z\mapsto az^{q^j}$ with $a\in L^*$ and $0\leq j<n$. They also observed that cyclic orbit codes with the same distance distribution need not be linearly equivalent. For $k=3$, Castello, Polverino and Zullo~\cite[Theorems~III.6 and~III.7]{CPZ24} determined when two spaces $U_3(\lambda)$ with $t_\lambda>4$, or two spaces $S_1(\lambda)$, are mapped onto each other by a map $z\mapsto az^{\sigma}$ with $a\in L^*$ and $\sigma\in\operatorname{Aut}(L)$.

The question of determining the linear-equivalence classes of the codes in Families~I and~II was posed as the second open problem in the concluding remarks of \cite{CPZ}. In general, the classification of subspace codes up to equivalence is a difficult problem; see \cite{HKK} and, for example, the classification of the optimal binary codes of length $8$, constant dimension $4$ and minimum distance $6$ in~\cite{HHKKW}. The two families are already known to be inequivalent by \cite[Corollary 4.12]{CPZ}. In this paper we determine the equivalence classes within each family for all $k\geq2$, and hence we solve this open problem. Our main result is the following.

\begin{theorem}\label{thm:main}
Fix $q,n,k$ with $k\geq2$.
\begin{enumerate}[label=\textup{(\roman*)}]
\item Let $\lambda,\mu\in L\setminus\F_q$ satisfy \eqref{eq:range-I}. The codes $\C_L(U_k(\lambda))$ and $\C_L(U_k(\mu))$ are linearly equivalent if and only if
\begin{equation}\label{eq:mobius}
 \mu=\frac{a\lambda^{q^j}+b}{c\lambda^{q^j}+d}
\end{equation}
for some $a,b,c,d\in\F_q$ with $ad-bc\neq0$ and some integer $0\leq j<n$.
\item Suppose $n$ is even, $K=\F_{q^2}$ and $k=2\ell+1$. Let $\lambda,\mu\in L\setminus K$ satisfy \eqref{eq:range-II}. The codes $\C_L(S_\ell(\lambda))$ and $\C_L(S_\ell(\mu))$ are linearly equivalent if and only if
\begin{equation}\label{eq:affine}
 \mu=a+b\lambda^{q^j}
\end{equation}
for some $a\in K$, $b\in K^*$ and some integer $0\leq j<n$.
\end{enumerate}
In each part, the degrees of the fields generated by $\lambda$ and $\mu$ over the indicated base field are then equal.
\end{theorem}

Since the two families are inequivalent and there is only one code for $k=1$, Theorem~\ref{thm:main} gives the linear-equivalence classes of all FWS one-orbit cyclic subspace codes.

\begin{remark}\label{rem:polynomials}
Let $m_\lambda\in\F_q[X]$ be the minimal polynomial of $\lambda$ over $\F_q$. Since fractional linear transformations over $\F_q$ commute with the $q$-Frobenius map, $\PGL_2(\F_q)$ acts on the monic irreducible polynomials over $\F_q$ of a fixed degree $t\geq2$ through its action on their roots; see~\cite{Gar,ST}. Part~(i) of Theorem~\ref{thm:main} says that $\C_L(U_k(\lambda))$ and $\C_L(U_k(\mu))$ are linearly equivalent if and only if $m_\mu$ lies in the $\PGL_2(\F_q)$-orbit of $m_\lambda$. In particular, the criterion does not depend on $k$. If $t$ is a prime not dividing $|\PGL_2(\F_q)|=q(q^2-1)$, then this action is free on the polynomials of degree $t$, since a nonidentity element fixing such a polynomial would act on its roots as a nontrivial power of the Frobenius map, and hence would have order $t$. Hence, for each admissible $k$, the codes $\C_L(U_k(\lambda))$ with $t_\lambda=t$ form exactly $(q^{t-1}-1)/(t(q^2-1))$ classes. For example, if $q=2$ and $t_\lambda=n=7$, then there are three classes for each $k\in\{2,3\}$, although for fixed $k$ these codes have the same distance distribution by \cite[Corollary~4.6]{CPZ}; for $k=3$ see also \cite[Remark~III.9]{CPZ24}. Part~(ii) can be stated in a similar way with minimal polynomials over $K$, using the substitutions $X\mapsto a+bX$ with $a\in K$, $b\in K^*$ and the nontrivial automorphism of $K$ acting on the coefficients.
\end{remark}

\begin{remark}\label{rem:semilinear}
Write $q=p^e$ with $p$ prime. Let $T$ be a semilinear bijection of $L$, that is, $T$ is additive and $T(cz)=c^{\sigma}T(z)$ for all $c\in\F_q$ and $z\in L$, where $\sigma$ is an automorphism of $\F_q$. Then $T$ is the composition of an $\F_q$-linear map with $z\mapsto z^{p^i}$ for some $i$. Since $z\mapsto z^{p^i}$ maps $\C_L(U_k(\lambda))$ onto $\C_L(U_k(\lambda^{p^i}))$ and $\C_L(S_\ell(\lambda))$ onto $\C_L(S_\ell(\lambda^{p^i}))$, Theorem~\ref{thm:main} also determines the equivalence of these codes under semilinear bijections if $\lambda^{q^j}$ with $0\leq j<n$ is replaced by $\lambda^{p^j}$ with $0\leq j<en$. For $k=3$ this agrees with~\cite{CPZ24}.
\end{remark}

The rest of this paper is organized as follows. In Section~\ref{sec:reduction}, we first reduce to the smallest field containing the representative subspace (Lemma~\ref{lem:field}). This permits us to apply Theorem~\ref{thm:generic}, the linear-isometry theorem of Gluesing-Luerssen and Lehmann~\cite{GLL}. We then determine the possible scalar equivalences by polynomial calculations (Lemmas~\ref{lem:scalar-I} and~\ref{lem:scalar-II}). For Family~I, the case $t_\lambda=2k-1$ is treated separately using the trace pairing (Lemma~\ref{lem:dual}). We prove part~(i) of Theorem~\ref{thm:main} in Section~\ref{sec:first} and part~(ii) in Section~\ref{sec:second}. Section~\ref{sec:conclusion} concludes the paper.

\section{Reduction to the generated field}\label{sec:reduction}

In this section we show that if two cyclic orbit codes are linearly equivalent, then their representatives containing $1$ generate the same subfield $E$ of $L$, and the equivalence can be restricted to $E$ (Lemma~\ref{lem:field}). Then we recall the linear-isometry theorem of Gluesing-Luerssen and Lehmann (Theorem~\ref{thm:generic}), which we apply in $E$.

For an $\F_q$-subspace $U\subseteq L$ containing $1$, let
\[
 E(U)=\F_q(U)
\]
be the smallest subfield of $L$ containing $U$. The degree of this field is invariant under linear equivalence; see \cite[Corollary 2.6]{CPZ}. The following argument also gives the restriction of an equivalence that we need.

\begin{lemma}\label{lem:field}
Let $1\in U\subseteq L$ and $\dim_{\F_q}U\geq2$. Form a graph whose vertices are the codewords of $\C_L(U)$, joining two distinct vertices when their intersection is nonzero. The set of distinct $\F_q$-linear spans of the connected components is
\[
 \{aE(U):a\in L^*\}.
\]
Consequently, if $1\in V$ and $\C_L(U)$ is linearly equivalent to $\C_L(V)$, then $E(U)=E(V)=E$. After multiplication by a scalar, the equivalence restricts to an $\F_q$-linear equivalence between $\C_E(U)$ and $\C_E(V)$.
\end{lemma}

\begin{proof}
Put $E=E(U)$, write $U^*=U\setminus\{0\}$, and let
\[
 H=\langle u/v:u,v\in U^*\rangle
\]
be the multiplicative subgroup of $E^*$ generated by these quotients. For $a,b\in L^*$,
\[
 aU\cap bU\neq\{0\}\quad\Longleftrightarrow\quad b/a\in U^*/U^*.
\]
It follows that the component containing $U$ consists of the subspaces $hU$ with $h\in H$. Since $1\in U$, we have $U^*\subseteq H$, and hence the span of this component is
\[
 \Span_{\F_q}\Bigl(\bigcup_{h\in H}hU\Bigr)=\Span_{\F_q}H.
\]
The space on the right contains $1$ and is closed under multiplication. It is a finite subring of a field, so it is a field. It contains $U$ and is contained in $E$, and therefore it equals $E$. Multiplication by any $a\in L^*$ gives the assertion for the component containing $aU$. Different components may have the same span, which is why we take the set of distinct spans.

An invertible $\F_q$-linear map preserves nonzero intersections, connected components and their spans. An equivalence therefore maps the set of scalar multiples of $E(U)$ onto the corresponding set for $E(V)$. These subspaces have the same dimension over $\F_q$. Since $L$ has a unique subfield of each possible order, $E(U)=E(V)=E$.

If the equivalence maps $E$ onto $bE$, multiply it by $b^{-1}$. Its restriction to $E$ is then an invertible $\F_q$-linear map of $E$. Finally, the codewords of $\C_L(U)$ contained in $E$ are exactly the codewords of $\C_E(U)$: if $aU\subseteq E$, then $a\in E$ because $1\in U$. The same holds for $V$, proving the last assertion.
\end{proof}

We use the following form of the linear-isometry theorem \cite[Theorem 6.2(a)]{GLL}; see also \cite[Theorem 2.3]{CPZ}.

\begin{theorem}[Gluesing-Luerssen--Lehmann]\label{thm:generic}
Let $E=\F_{q^s}$ and let $r$ be an integer with $2\leq r\leq s/2$. Let $U,V$ be $r$-dimensional $\F_q$-subspaces of $E$, with $1\in V$ and $\F_q(V)=E$. Then $\C_E(U)$ and $\C_E(V)$ are linearly equivalent if and only if
\[
 V=aU^{q^j}
 \quad\text{for some }a\in E^*\text{ and }0\leq j<s,
\]
where $U^{q^j}=\{u^{q^j}:u\in U\}$.
\end{theorem}

The bound $2r\leq s$ is needed here. In our applications, Lemma~\ref{lem:field} first replaces $L$ by the field generated by the subspace. We then check this dimension bound in that field.

\section{Polynomial spaces and the first family}\label{sec:first}

In this section we determine when two polynomial spaces are scalar multiples of each other (Lemma~\ref{lem:scalar-I}), and we prove part~(i) of Theorem~\ref{thm:main}. For the case $t_\lambda=2k-1$ we also use the trace dual of a polynomial space given in Lemma~\ref{lem:dual}.

Let $B$ be a finite field and let $M$ be a finite extension of $B$. For $x\in M$, put $t=[B(x):B]$. For an integer $1\leq r\leq t$, write
\[
 P_r(x)=\Span_B\{1,x,\ldots,x^{r-1}\}\subseteq M.
\]
Evaluation at $x$ is injective on polynomials over $B$ of degree less than $t$. The next lemma is the polynomial-intersection calculation used in \cite[Lemma 4.3]{CPZ}. We give the proof in the range $t>2r-2$, which will also cover the boundary case below.

\begin{lemma}\label{lem:intersection}
Let $x\in M$ have degree $t$ over $B$, and let $r\geq2$ with $t>2r-2$. Let $f,g\in B[X]$ be nonzero coprime polynomials of degrees at most $r-1$. Put $z=f(x)/g(x)$. Then
\[
 \dim_B\bigl(P_r(x)\cap zP_r(x)\bigr)
 =r-\max\{\deg f,\deg g\}.
\]
\end{lemma}

\begin{proof}
An element of the intersection has the form $A(x)=zD(x)$, where $A,D\in B[X]$ have degrees at most $r-1$. Thus
\[
 (gA-fD)(x)=0.
\]
Since $\deg(gA-fD)\leq2r-2<t$, this gives the polynomial identity $gA=fD$. As $f$ and $g$ are coprime, $A=fh$ and $D=gh$ for some $h\in B[X]$. Conversely, every such pair gives an element of the intersection. The permitted polynomials $h$ have degree less than $r-\max\{\deg f,\deg g\}$. Evaluation and multiplication by $f(x)\neq0$ are injective on this space, giving the stated dimension.
\end{proof}

\begin{lemma}\label{lem:scalar-I}
Let $r\geq2$ and let $x,y\in M$ have degrees at least $2r-1$ over $B$. Then $P_r(y)=\alpha P_r(x)$ for some $\alpha\in M^*$ if and only if
\[
 y=\frac{ax+b}{cx+d}
 \quad\text{for some }a,b,c,d\in B\text{ with }ad-bc\neq0.
\]
\end{lemma}

\begin{proof}
Suppose $P_r(y)=\alpha P_r(x)$. Since $\alpha,\alpha x\in P_r(y)$, we may write $x=f(y)/g(y)$ with $f,g\in B[X]$ nonzero and coprime, both of degree at most $r-1$. The powers $1,x,\ldots,x^r$ are linearly independent over $B$, so
\[
 \dim_B\bigl(P_r(x)\cap xP_r(x)\bigr)=r-1.
\]
Multiplication by $\alpha$ and Lemma~\ref{lem:intersection} now give
\[
 r-1=\dim_B\bigl(P_r(y)\cap xP_r(y)\bigr)
 =r-\max\{\deg f,\deg g\}.
\]
Thus $\max\{\deg f,\deg g\}=1$. The quotient is nonconstant because $x\notin B$. Its coefficient matrix is therefore invertible, and inversion gives the required expression for $y$.

Conversely, suppose $y=(ax+b)/(cx+d)$ with $ad-bc\neq0$. The denominator is nonzero since $x\notin B$. For $0\leq i<r$,
\[
 (cx+d)^{r-1}y^i=(ax+b)^i(cx+d)^{r-1-i}\in P_r(x).
\]
Both spaces have dimension $r$, and hence
\begin{equation}\label{eq:scalar-factor}
 (cx+d)^{r-1}P_r(y)=P_r(x).
\end{equation}
This proves the converse.
\end{proof}

For the case $t_\lambda=2k-1$, we shall also need the following trace-dual identity; see \cite[Proposition 3.3]{CPZ}.

\begin{lemma}\label{lem:dual}
Let $x\in E=\F_{q^t}$ satisfy $[\F_q(x):\F_q]=t$, and let $f\in\F_q[X]$ be its monic minimal polynomial. For $1\leq k<t$, put $U_k(x)=\Span_{\F_q}\{1,x,\ldots,x^{k-1}\}$. With respect to the trace pairing
\[
 (u,v)\longmapsto\Tr_{E/\F_q}(uv),
\]
we have
\begin{equation}\label{eq:dual}
 U_k(x)^\perp=f'(x)^{-1}U_{t-k}(x).
\end{equation}
Here $f'$ is the formal derivative of $f$, and $f'(x)\neq0$.
\end{lemma}

\begin{proof}
The polynomial $f$ divides $X^{q^t}-X$, whose derivative is $-1$. Thus $f$ is separable and $f'(x)\neq0$.

For $0\leq i\leq t-2$, Lagrange interpolation at the roots of $f$ gives
\[
 X^i=\sum_{f(\alpha)=0}\alpha^i\,
 \frac{f(X)}{(X-\alpha)f'(\alpha)}.
\]
Comparing coefficients of $X^{t-1}$ yields
\[
 \Tr_{E/\F_q}\bigl(x^i/f'(x)\bigr)
 =\sum_{f(\alpha)=0}\frac{\alpha^i}{f'(\alpha)}=0.
\]
Products of a basis element of $U_k(x)$ and a basis element of $U_{t-k}(x)$ have the form $x^i$ with $0\leq i\leq t-2$. Therefore $f'(x)^{-1}U_{t-k}(x)\subseteq U_k(x)^\perp$. The trace pairing is nondegenerate, and both sides have dimension $t-k$. This proves \eqref{eq:dual}.
\end{proof}

\begin{proof}[Proof of Theorem~\ref{thm:main}\textup{(i)}]
Suppose the two codes are linearly equivalent. Their representatives contain $1$ and generate $\F_q(\lambda)$ and $\F_q(\mu)$, respectively. By Lemma~\ref{lem:field}, these fields are equal, say $E=\F_{q^t}$, and the equivalence restricts to the orbit codes in $E$.

First suppose $2k\leq t$. Theorem~\ref{thm:generic} gives
\[
 U_k(\mu)=\alpha U_k(\lambda)^{q^j}
 =\alpha U_k(\lambda^{q^j})
\]
for some $\alpha\in E^*$ and $0\leq j<t$. Lemma~\ref{lem:scalar-I}, with $B=\F_q$, now gives \eqref{eq:mobius}.

It remains to consider $t=2k-1<n$. Suppose first that $k\geq3$. Let $T$ be an $\F_q$-linear equivalence between $\C_E(U_k(\lambda))$ and $\C_E(U_k(\mu))$. Define its adjoint $T^*$ by
\[
 \Tr_{E/\F_q}(T(u)v)=\Tr_{E/\F_q}(uT^*(v)).
\]
For every subspace $W\subseteq E$,
\[
 T(W)^\perp=(T^*)^{-1}(W^\perp).
\]
Thus $(T^*)^{-1}$ maps the dual codes onto one another. Since $(aW)^\perp=a^{-1}W^\perp$, Lemma~\ref{lem:dual} identifies these dual orbit codes with
\[
 \C_E(U_{k-1}(\lambda))
 \quad\text{and}\quad
 \C_E(U_{k-1}(\mu)).
\]
Both representatives still contain $1$ and the corresponding parameter. Moreover, $2\leq k-1\leq t/2$. Applying Theorem~\ref{thm:generic} and Lemma~\ref{lem:scalar-I} to these spaces again gives \eqref{eq:mobius}.

If $k=2$ and $t=3$, the group $\PGL_2(\F_q)$ acts freely on the degree-three elements of $\F_{q^3}$. Indeed, a nonidentity fractional linear transformation fixing such an element would give a nonzero polynomial of degree at most two having it as a root. There are $q^3-q$ degree-three elements, and $|\PGL_2(\F_q)|=q^3-q$. The action is therefore transitive, so any two such parameters satisfy \eqref{eq:mobius} with $j=0$.

For the converse, $z\mapsto z^{q^j}$ is an invertible $\F_q$-linear map on $L$ and sends $\C_L(U_k(\lambda))$ onto $\C_L(U_k(\lambda^{q^j}))$. Equation~\eqref{eq:scalar-factor} shows that the latter orbit code equals $\C_L(U_k(\mu))$. This proves the assertion.
\end{proof}

\section{The second family}\label{sec:second}

In this section we prove part~(ii) of Theorem~\ref{thm:main}. We show that $S_\ell(\lambda)$ determines the space $K+K\lambda$ (Lemma~\ref{lem:second-spaces}), and using this we determine when $S_\ell(\mu)$ is a scalar multiple of $S_\ell(\lambda)$ (Lemma~\ref{lem:scalar-II}).

Throughout this section, $n$ is even, $K=\F_{q^2}\subseteq L$, and $\ell\geq1$. Let $\lambda\in L\setminus K$ with $t=[K(\lambda):K]>2\ell$. To simplify the notation, put
\begin{align*}
 A&=\Span_K\{1,\lambda,\ldots,\lambda^{\ell-1}\},\\
 S&=A\oplus\F_q\lambda^\ell,\\
 B&=\Span_K\{1,\lambda,\ldots,\lambda^\ell\}.
\end{align*}
The next lemma shows that $A$ is the largest $K$-linear subspace contained in $S$, and that $B$ is the $K$-linear span of $S$, that is, the space of all finite sums of elements $as$ with $a\in K$ and $s\in S$.

\begin{lemma}\label{lem:second-spaces}
For any $\omega\in K\setminus\F_q$,
\[
 S\cap\omega S=A,\qquad \Span_K S=B.
\]
Every $K$-linear subspace of $S$ is contained in $A$. Furthermore,
\begin{equation}\label{eq:multiplication-space}
 \{z\in L:zA\subseteq B\}=K+K\lambda.
\end{equation}
\end{lemma}

\begin{proof}
The elements $1,\lambda,\ldots,\lambda^\ell$ are linearly independent over $K$, and multiplication by $\omega$ preserves $A$. Consequently,
\[
 S\cap\omega S
 =A\oplus(\F_q\cap\omega\F_q)\lambda^\ell=A.
\]
If $W\subseteq S$ is $K$-linear, then $W=\omega W\subseteq\omega S$, and therefore $W\subseteq A$. Taking the $K$-span of $S$ gives $B$.

For \eqref{eq:multiplication-space}, suppose first that $\ell=1$. Then $A=K$ and $B=K+K\lambda$, so $zA\subseteq B$ is equivalent to $z\in B$.

Now let $\ell\geq2$ and suppose $zA\subseteq B$. Since $1\in A$, we have $z\in B$, so $z=h(\lambda)$ for a polynomial $h\in K[X]$ of degree at most $\ell$. If $h=0$, the assertion is immediate. Otherwise, $\lambda^{\ell-1}\in A$ implies
\[
 h(\lambda)\lambda^{\ell-1}\in B.
\]
The polynomial $h(X)X^{\ell-1}$ has degree at most $2\ell-1<t$. Its value at $\lambda$ belongs to $B$ only if this polynomial has degree at most $\ell$, by injectivity of evaluation below degree $t$. Hence $\deg h\leq1$, giving $z\in K+K\lambda$. Conversely, each element of $K+K\lambda$ maps $A$ into $B$ by multiplication.
\end{proof}

For any $c\in L^*$ we have
\begin{equation}\label{eq:cancel}
 \{z\in L:z(cA)\subseteq cB\}
 =\{z\in L:zA\subseteq B\},
\end{equation}
since $c$ can be cancelled. This gives the following scalar-equivalence criterion.

\begin{lemma}\label{lem:scalar-II}
Let $\lambda,\mu\in L\setminus K$ satisfy $[K(\lambda):K]>2\ell$ and $[K(\mu):K]>2\ell$. Then $S_\ell(\mu)=cS_\ell(\lambda)$ for some $c\in L^*$ if and only if
\[
 \mu=a+b\lambda,\qquad a\in K,\quad b\in K^*.
\]
\end{lemma}

\begin{proof}
Suppose $S_\ell(\mu)=cS_\ell(\lambda)$. Multiplication by $c$ is $K$-linear. It therefore carries the largest $K$-subspace of $S_\ell(\lambda)$ onto that of $S_\ell(\mu)$, and carries their $K$-spans onto one another. By Lemma~\ref{lem:second-spaces},
\[
 A_\ell(\mu)=cA_\ell(\lambda),\qquad
 A_{\ell+1}(\mu)=cA_{\ell+1}(\lambda).
\]
Equations~\eqref{eq:multiplication-space} and \eqref{eq:cancel} imply
\[
 K+K\mu=K+K\lambda.
\]
Thus $\mu=a+b\lambda$ for $a,b\in K$, with $b\neq0$ since $\mu\notin K$.

Conversely, let $\mu=a+b\lambda$ with $a\in K$ and $b\in K^*$. Expanding powers gives
\[
 A_\ell(\mu)=A_\ell(\lambda),\qquad
 \mu^\ell-b^\ell\lambda^\ell\in A_\ell(\lambda).
\]
Indeed, the change between the two polynomial bases is triangular with nonzero diagonal entries $1,b,\ldots,b^{\ell-1}$. Therefore
\begin{align*}
 S_\ell(\mu)
 &=A_\ell(\lambda)\oplus\F_q b^\ell\lambda^\ell\\
 &=b^\ell S_\ell(\lambda),
\end{align*}
as required.
\end{proof}

\begin{proof}[Proof of Theorem~\ref{thm:main}\textup{(ii)}]
The smallest field containing $S_\ell(\lambda)$ is $K(\lambda)$. Indeed, $K\subseteq S_\ell(\lambda)$ and $\lambda\in S_\ell(\lambda)$, also when $\ell=1$. An ambient linear equivalence therefore restricts, by Lemma~\ref{lem:field}, to an equivalence of the orbit codes in
\[
 E=K(\lambda)=K(\mu)=\F_{q^{2t}}.
\]
Since $t>2\ell$, the codeword dimension satisfies $2\ell+1\leq t$. Thus Theorem~\ref{thm:generic} applies with $r=2\ell+1$ and $s=2t$, and gives
\[
 S_\ell(\mu)=cS_\ell(\lambda)^{q^j}
 =cS_\ell(\lambda^{q^j})
\]
for some $c\in E^*$ and $0\leq j<2t$. The last equality holds because $q$-Frobenius preserves $K$ and fixes $\F_q$. Lemma~\ref{lem:scalar-II} now gives \eqref{eq:affine}.

Conversely, if \eqref{eq:affine} holds, then
\[
 S_\ell(\mu)=b^\ell S_\ell(\lambda^{q^j}).
\]
The map $z\mapsto z^{q^j}$ on $L$ sends the first orbit code onto the second. This proves the theorem.
\end{proof}

\section{Conclusion}\label{sec:conclusion}

In this paper we determined the linear-equivalence classes of all FWS one-orbit cyclic subspace codes, and hence solved the second open problem posed by Castello, Polverino and Zullo in~\cite{CPZ}. In Family~I, two codes are linearly equivalent if and only if their parameters $\lambda$ and $\mu$ are related by a fractional linear transformation over $\F_q$ and a power of $q$-Frobenius. In Family~II, the corresponding condition is an affine transformation over $\F_{q^2}$ and a power of $q$-Frobenius.

\section*{Acknowledgements}

The research of M. Shi was supported by the National Natural Science Foundation of China under Grants W2611002 and 12471490.


\begin{thebibliography}{99}
\bibitem{Al}
T. L. Alderson,
\emph{A note on full weight spectrum codes},
Trans. Comb. \textbf{8} (2019), no. 3, 15--22.

\bibitem{AN}
T. L. Alderson and A. Neri,
\emph{Maximum weight spectrum codes},
Adv. Math. Commun. \textbf{13} (2019), no. 1, 101--119.

\bibitem{BEGR}
E. Ben-Sasson, T. Etzion, A. Gabizon and N. Raviv,
\emph{Subspace polynomials and cyclic subspace codes},
IEEE Trans. Inform. Theory \textbf{62} (2016), no. 3, 1157--1165.

\bibitem{CPSZ}
C. Castello, O. Polverino, P. Santonastaso and F. Zullo,
\emph{Constructions and equivalence of Sidon spaces},
J. Algebraic Combin. \textbf{58} (2023), no. 4, 1299--1329.

\bibitem{CPZ24}
C. Castello, O. Polverino and F. Zullo,
\emph{On one-orbit cyclic subspace codes of $\mathcal{G}_q(n,3)$},
in: Proc. 2024 IEEE Int. Symp. Inf. Theory (ISIT), Athens, 2024, pp. 2383--2388.

\bibitem{CPZ}
C. Castello, O. Polverino and F. Zullo,
\emph{Full weight spectrum one-orbit cyclic subspace codes},
J. Combin. Theory Ser. A \textbf{212} (2025), 106005.

\bibitem{De73}
Ph. Delsarte,
\emph{Four fundamental parameters of a code and their combinatorial significance},
Information and Control \textbf{23} (1973), no. 5, 407--438.

\bibitem{EV}
T. Etzion and A. Vardy,
\emph{Error-correcting codes in projective space},
IEEE Trans. Inform. Theory \textbf{57} (2011), no. 2, 1165--1173.

\bibitem{Gar}
T. Garefalakis,
\emph{On the action of $\GL_2(\F_q)$ on irreducible polynomials over $\F_q$},
J. Pure Appl. Algebra \textbf{215} (2011), no. 8, 1835--1843.

\bibitem{GL}
H. Gluesing-Luerssen and H. Lehmann,
\emph{Distance distributions of cyclic orbit codes},
Des. Codes Cryptogr. \textbf{89} (2021), no. 3, 447--470.

\bibitem{GLL}
H. Gluesing-Luerssen and H. Lehmann,
\emph{Automorphism groups and isometries for cyclic orbit codes},
Adv. Math. Commun. \textbf{17} (2023), no. 1, 119--138.

\bibitem{HHKKW}
D. Heinlein, T. Honold, M. Kiermaier, S. Kurz and A. Wassermann,
\emph{Classifying optimal binary subspace codes of length $8$, constant dimension $4$ and minimum distance $6$},
Des. Codes Cryptogr. \textbf{87} (2019), no. 2--3, 375--391.

\bibitem{HKK}
T. Honold, M. Kiermaier and S. Kurz,
\emph{Constructions and bounds for mixed-dimension subspace codes},
Adv. Math. Commun. \textbf{10} (2016), no. 3, 649--682.

\bibitem{KK}
R. Koetter and F. R. Kschischang,
\emph{Coding for errors and erasures in random network coding},
IEEE Trans. Inform. Theory \textbf{54} (2008), no. 8, 3579--3591.

\bibitem{OO1}
K. Otal and F. \"Ozbudak,
\emph{Cyclic subspace codes via subspace polynomials},
Des. Codes Cryptogr. \textbf{85} (2017), no. 2, 191--204.

\bibitem{OO2}
K. Otal and F. \"Ozbudak,
\emph{Constructions of cyclic subspace codes and maximum rank distance codes},
in: Network Coding and Subspace Designs, Signals Commun. Technol., Springer, Cham, 2018, pp. 43--66.

\bibitem{OSZ}
F. \"Ozbudak, P. Santonastaso and F. Zullo,
\emph{Using multi-orbit cyclic subspace codes for constructing optical orthogonal codes},
Cryptogr. Commun. \textbf{17} (2025), no. 4, 695--708.

\bibitem{RRT}
R. M. Roth, N. Raviv and I. Tamo,
\emph{Construction of Sidon spaces with applications to coding},
IEEE Trans. Inform. Theory \textbf{64} (2018), no. 6, 4412--4422.

\bibitem{SLNS}
M. Shi, X. Li, A. Neri and P. Sol\'e,
\emph{How many weights can a cyclic code have?},
IEEE Trans. Inform. Theory \textbf{66} (2020), no. 3, 1449--1459.

\bibitem{SZSC}
M. Shi, H. Zhu, P. Sol\'e and G. D. Cohen,
\emph{How many weights can a linear code have?},
Des. Codes Cryptogr. \textbf{87} (2019), no. 1, 87--95.

\bibitem{ST}
H. Stichtenoth and A. Topuzo\u{g}lu,
\emph{Factorization of a class of polynomials over finite fields},
Finite Fields Appl. \textbf{18} (2012), no. 1, 108--122.

\bibitem{Tr}
A.-L. Trautmann,
\emph{Isometry and automorphisms of constant dimension codes},
Adv. Math. Commun. \textbf{7} (2013), no. 2, 147--160.

\bibitem{TMBR}
A.-L. Trautmann, F. Manganiello, M. Braun and J. Rosenthal,
\emph{Cyclic orbit codes},
IEEE Trans. Inform. Theory \textbf{59} (2013), no. 11, 7386--7404.

\bibitem{Zu}
F. Zullo,
\emph{Multi-orbit cyclic subspace codes and linear sets},
Finite Fields Appl. \textbf{87} (2023), 102153.
\end{thebibliography}
\end{document}